\documentclass[12pt]{article}
 \usepackage{amsmath,amsfonts,latexsym,amssymb,amsthm}
 \usepackage{times}
 \usepackage{amscd}

   \newcommand{\field}[1]{\mathbb{#1}}
   \newcommand{\rz}{\field{R}}
   \newcommand{\cz}{\field{C}}
   \newcommand{\zz}{\field{Z}}
   \newcommand{\nz}{\field{N}}

   \newcommand{\co}{C^\infty _0}

   \newcommand{\Hilbert}{\mathcal{H}}
   
   \newcommand{\alg}{\mathcal{A}}

   \newcommand{\Spin}{\mathrm{Spin}}
   \newcommand{\SO}{\mathrm{SO}}
   \newcommand{\Cl}{\mathrm{Cl}}
   \newcommand{\Clc}{\mathrm{Cl}^c}
   \newcommand{\Mat}{\mathrm{Mat}_\cz}

   \newcommand{\Dix}{\mathrm{Tr}_\omega}
   \newcommand{\dom}{\mathrm{dom}}
   \newcommand{\Di}{\mathcal{D}'}
   \newcommand{\op}{\mathrm{op}}
   
   \newcommand{\I}{\mathrm{i}}
   \newcommand{\Tr}{\mathrm{Tr}}
   \newcommand{\J}{\mathfrak{J}}
   \newcommand{\Dirac}{/\!\!\!\!D}

 \newtheorem{theorem}{Theorem}[section]
 \newtheorem{definition}[theorem]{Definition}
 \newtheorem{exa}[theorem]{Example}
 \newtheorem{lem}[theorem]{Lemma}
 
 \newtheorem{pro}[theorem]{Proposition}

\begin{document}

\title{On Noncommutative and semi-Riemannian Geometry}

\author{Alexander Strohmaier}

\date{\small Universit\"at Leipzig,
Institut f\"ur theoretische Physik,
Augustusplatz 10/11, D-04109 Leipzig, Germany\\
E-mail: alexander.strohmaier@itp.uni-leipzig.de}
\maketitle
\noindent
\begin{abstract}
  \noindent
  We introduce the notion of a semi-Riemannian spectral triple which generalizes
  the notion of spectral triple and allows for a treatment of
  semi-Riemannian manifolds within a noncommutative setting.
  It turns out that the relevant spaces in noncommutative semi-Riemannian
  geometry are not Hilbert spaces any more but Krein spaces, and Dirac
  operators are Krein-selfadjoint. We show that the noncommutative tori
  can be endowed with a semi-Riemannian structure in this way. For the
  noncommutative tori as well as for semi-Riemannian spin manifolds
  the dimension, the signature of the metric, and the integral of a function
  can be recovered from the spectral data.
\end{abstract}

{\small \bf Mathematics Subject Classification (2000):} 58B34, 58B99, 46C20,
53C50 \\
\maketitle

\section{Introduction}

The Gel'fand-Naimark theorem states that any unital commutative $C^*$-algebra
can be realized as an algebra of continuous functions on a compact Hausdorff
space. In noncommutative geometry one thinks of a noncommutative
$C^*$-algebra as an algebra of functions on some "virtual" space and tries to
imitate geometrical constructions which work for the case of commutative
algebras.
Connes functional analytic approach (see \cite{Connes:1994gy}) to noncommutative geometry starts with
the observation that the metric information of a compact Riemannian spin manifold $M$
is encoded in the triple $(C^\infty(M),\Dirac,L^2(M,S))$, where $\Dirac$
is the Dirac operator and $L^2(M,S)$ is the Hilbert space of square
integrable sections of the spinor bundle. The algebra $C^\infty(M)$ is
realized as a $*$-algebra of bounded operators on $L^2(M,S)$.
The space of characters of $C^\infty(M)$
is canonically isomorphic to the set of points of $M$ and the Riemannian
distance between to point $p$ and $q$ can be recovered from the equation
\begin{gather}
   d(p,q)=\sup \vert f(p)-f(q) \vert;\; f \in C^\infty(M),\; \Vert [\Dirac,f] \Vert \leq 1.
\end{gather}
The noncommutative generalization of the object
$(C^\infty(M),\Dirac,L^2(M,S))$
is the so called spectral triple, which we
can think of as a generalization of the theory of compact Riemannian
manifolds.
For a further introduction to noncommutative geometry we would like to refer
the reader to \cite{Gracia-Bondia:2001tr}, \cite{Landi:1997}, \cite{Varilly},
\cite{Madore:2000fg}
and the references therein.

Recently there have been attempts to get analogues of spectral triples which
allow for a treatment of non-compact manifolds (see \cite{Rennie:00sd}) and of
globally hyperbolic Lorentzian manifolds
(see \cite{Parfionov:1998xs,Kopf:2001fq,Kopf:2001de,Kopf:2000fq}).
Such a treatment seems necessary if one wants to study physical models, which
are defined on spaces with Lorentzian rather than Riemannian metrics.
The idea in \cite{Hawk:97},\cite{Kopf:2001fq}, \cite{Kopf:2001de} and \cite{Kopf:2000fq} is to foliate
the spacetime into Cauchy surfaces and to treat the Cauchy surfaces
as Riemannian manifolds. Whereas this approach seems promising for the study of
evolution equations in physics, its dependence on the foliation and the
restriction to Lorentzian signatures is disturbing from the
mathematical point of view. 

In this paper we suggest a notion of semi-Riemannian spectral triple, which
allows to treat compact semi-Riemannian manifolds (of arbitrary signature)
within noncommutative geometry. Such a triple $(\alg,D,\Hilbert)$ consists of
an involutive algebra $\alg$ of bounded operators acting on a Krein space
$\Hilbert$ and a Krein-selfadjoint operator $D$. An important role is played
by the fundamental symmetries of the Krein space. These are operators $\J :
\Hilbert \to \Hilbert$ with $\J^2=1$ such that $(\cdot,\J\cdot)=(\J \cdot,\cdot)$ is a positive definite
scalar product turning $\Hilbert$ into a Hilbert space. They can be used to
obtain ordinary spectral triples from semi-Riemannian spectral triples
in a similar way as this is done in physics by "Wick rotation",
which is used to pass to Riemannian signatures of the metric. For example
if $M$ is a Lorentzian spin manifold, $\alg=C_0^\infty(M)$ and $D$ is the Dirac operator which acts
on the Krein space of square integrable sections $\Hilbert$ of the spinor
bundle, the triple $(\alg,D,\Hilbert)$ is a semi-Riemannian spectral triple.
For a special class of fundamental symmetries the operator
$\frac{1}{2}((\J D)^2 + (D \J)^2)$ is a Laplace-type operator with respect to a Riemannian
metric. We can think of this metric as a Wick rotated form of the Lorentzian
metric.
We use this to show that one can define a notion of dimension for
semi-Riemannian spectral triples. In the commutative case and for the
noncommutative semi-Riemannian torus we show that there is a canonical
notion of integration and one can recover the signature of the metric from
the spectral data.

In sections 2-5 we review the basic notions and results on spectral
triples, Krein spaces and Dirac operators on semi-Riemannian manifolds.
Sections 6 and 7 contain the main results of this paper. 

\section{Spectral triples}

\begin{definition}
  A spectral triple $(\alg,\Hilbert,D)$ consists of a unital $*$-algebra
  $\alg$ of bounded operators on a separable Hilbert space $\Hilbert$
  and a selfadjoint operator $D$ on $\Hilbert$, such that the
  commutator $[D,a]$ is bounded for all $a \in \alg$.
  A spectral triple is said to be even if there exists an operator
  $\chi=\chi^*, \chi^2=1$ on the Hilbert space 
  such that
  \begin{gather}
    \chi a = a \chi, \quad \forall a \in \alg,\\
    D \chi = - \chi D.
  \end{gather}
\end{definition}

For a compact operator $a$ denote by $\mu_k(a)$ the ordered sequence of its
singular values, i.e. $\mu_k(a)$ are the eigenvalues of $\vert a \vert$
such that $\mu_1(a) \geq \mu_2(a) \geq \ldots \;$, with each eigenvalue repeated
according to its multiplicity.
The characteristic sequence of $a$ is defined by $\sigma_k(a):=\sum_{i=1}^k
\mu_i(a)$.
Let $p \geq 1$ be a real number.
A compact operator $a$ is said to be in $\mathcal{L}^{p+}$
if
\begin{gather}
 \sup_{n \geq 1} \frac{\sigma_n(a)}{n^{(p-1)/p}} < \infty \quad \textrm{ for } \quad p > 1,\\
 \sup_{n > 2} \frac{\sigma_n(a)}{\ln n} < \infty  \quad \textrm{ for
 } \quad p=1.
\end{gather}
The spaces $\mathcal{L}^{p+}$ are two 2-sided ideals in $\mathcal{B}(\Hilbert)$.
Note that if $a \in \mathcal{L}^{p+}$, then $\vert a \vert^p \in
\mathcal{L}^{1+}$.

Let now $l^\infty(\nz)$ be the von Neumann algebra of  bounded sequences.
If a state $\omega$ on $l^\infty(\nz)$ satisfies the conditions
\begin{itemize}
 \item $\lim_{n \to \infty} x_n = x \quad \Rightarrow \quad
   \omega((x_n))=x$,
 \item $\omega((x_{2n}))=\omega((x_n))$,
\end{itemize}
we say that $\omega$ is in $\Gamma_s(l^\infty)$. The set $\Gamma_s(l^\infty)$
turns out to be non-empty (\cite{Dixmier:1966df}). 
For each positive $a \in \mathcal{L}^{1+}$ and each state $\omega \in \Gamma_s(l^\infty)$ we define
$\Dix(a):=\omega(x(a))$, where $x(a)_n =
\frac{\sigma_n(a)}{\ln n}$ for $n \geq 2$ and $x(a)_1=0$.
It can be shown that for each $\omega$ the map $a \to \Dix(a)$ extends
to a finite trace on $\mathcal{L}^{1+}$ and to a singular trace on
$\mathcal{B}(\Hilbert)$ (see \cite{Dixmier:1966df, Connes:1994gy}).

\begin{definition}
 Let $p \geq 1$ be a real number.
 A spectral triple is called $p^+$-summable if $(1+D^2)^{-1/2}$
 is in $\mathcal{L}^{p+}$.
\end{definition}

In case a spectral triple is $p^+$-summable the map $a \to \Dix(a
(1+D^2)^{-p/2})$ is well defined on the algebra $\alg_D$ generated by $\alg$
and $[D,\alg]$. It can be shown that if $\alg_D$
is contained in the domain of smoothness of the derivation $\delta(\cdot):=[\vert D
\vert,\cdot]$, this map is a trace (see \cite{Cipr:1996df}).
A differential operator on a Riemannian manifold is said
to be of Dirac type if it is of first order and the principal symbol $\sigma$ of $D$ satisfies the
relation
\begin{gather}
 \sigma(\xi)^2 = g(\xi,\xi) \; \textrm{id}_{E_p} \quad \forall \xi \in T_p^*M, p \in M.
\end{gather}
The geometry of a compact Riemannian spin manifold can be encoded in a
spectral triple (see \cite{Connes:1994gy}).

\begin{theorem}[Connes] \label{class}
 Let $M$ be a compact Riemannian manifold of dimension $n$ and $E$ a hermitian vector bundle
 over $M$ of rank $k$. Let $\Hilbert$ be the Hilbert space of square
 integrable sections of $E$
 and let $\alg$ be $C^\infty(M)$ which acts on $\Hilbert$ by multiplication.
 Assume that $D$ is a symmetric differential operator of Dirac type on $E$.
 Then $(\alg,\Hilbert,D)$ is an $n^+$-summable spectral triple.
 As a compact space $M$ is the spectrum of the $C^*$-algebra, which is the norm closure
 of $\mathcal{A}$.
 The geodesic distance on $M$ is given by
 \begin{gather}
   d(p,q)=\sup \vert f(p)-f(q) \vert;\; f \in \alg,\; \Vert [D,f] \Vert \leq 1.
 \end{gather}
 Furthermore for $f \in C^\infty(M)$ we have
 \begin{gather}
  \int_M f \sqrt{\vert g \vert} dx = c(n,k) \Dix(f (1+\vert D \vert^2)^{-n/2}),
 \end{gather}
 where $c(n,k)=2^{n-1} \pi^{n/2} k^{-1} n \Gamma(n/2)$.
\end{theorem}

\section{Differential calculus and spectral triples}

Let $\alg$ be a unital algebra. 
Denote by $\overline \alg$ the vector space $\alg / (\cz 1)$ and define
$\Omega^n\alg:=\alg \otimes \overline \alg ^{\otimes n}$.
We write $(a_0,a_1,\ldots,a_n)$ for the image of $a_0 \otimes \ldots
\otimes a_n$ in $\Omega^n \alg$. On
$\Omega \alg:= \oplus_{n=0}^{\infty} \Omega^n \alg$ one now defines an
operator $d$ of degree one and a product by
\begin{gather}
  d(a_0,\ldots,a_n)=(1,a_0,\ldots,a_n),\\
  (a_0,\ldots,a_n)(a_{n+1},\ldots,a_{k})= \sum_{i=0}^n (-1)^{n-i}
  (a_0,\ldots,a_i a_{i+1},\ldots,a_k).
\end{gather}
This determines a differential algebra structure on $\Omega \alg$.
If $\alg$ is a star algebra, one makes $\Omega \alg$ a star algebra by
$(a_0,\ldots,a_n)^*:=(-1)^n (a_n^*,\ldots,a_1^*) \cdot a_0^*$.
The pair $(\Omega \alg,d)$ is called the
universal differential envelope of $\alg$.

A spectral triple $(\alg,\Hilbert,D)$ gives rise to a $*$-representation
of $\Omega \alg$ on $\Hilbert$ by the map
\begin{gather*}
 \pi : \Omega \alg \to \mathcal{B}(\Hilbert),\\
 \pi((a_0,a_1,\ldots,a_n)):= a_0 [D,a_1] \cdots [D,a_n], \quad a_j \in \alg.
\end{gather*}
Let $j_0$ be the graded two-sided ideal $j_0 := \oplus_{n} j_0^n$ given by
\begin{gather}
 j_0^n := \{ \omega \in \Omega^n \alg ; \pi(\omega)=0 \}.
\end{gather}
In general, $j_0$ is not a differential ideal. That is why it is 
not possible to define the space of forms to be the image
$\pi(\Omega \alg)$.
However $j:=j_0 + d j_0$ is a graded differential two-sided ideal.
\begin{definition}
 The graded differential algebra of Connes´ forms over $\alg$ is defined by
 \begin{gather}
  \Omega_D \alg := \Omega\alg / j \cong \bigoplus_{n} \pi(\Omega^n \alg)/
  \pi(d j_0 \cap \Omega^n \alg).
 \end{gather}
\end{definition}

\begin{exa}
 The space of one-forms $\Omega_D^1 \alg \cong \pi(\Omega^1 \alg)$
 is the space of bounded operators of the form
 \begin{gather}
  \omega_1=\sum_k a_0^k [D,a_1^k], \quad a_i^k \in \alg.
 \end{gather}
\end{exa}

\begin{pro} \label{bla}
 Let $(\alg,\Hilbert,D)$ be as in Theorem \ref{class}.
 As graded differential algebras $\Omega_D \alg$ and $\Gamma(\Lambda M)$
 are isomorphic.
\end{pro}\noindent
See \cite{Connes:1994gy}, p.552 or \cite{Landi:1997}, section 7.2.1 for a proof.\\
If the spectral triple $(\alg,\Hilbert,D)$ is $n^+$-summable the map
\begin{gather}
 w_1 \times w_2 \to \langle w_1, w_2 \rangle:= \Dix(w_1^* w_2 (1+\vert D \vert^2)^{-n/2})
\end{gather}
defines for each $\omega$ a scalar product on the space of one-forms.
In the case of Proposition \ref{bla} this scalar product coincides up to a scalar
factor with the metric-induced scalar product on the space of one-forms.

\section{Krein Spaces}

\subsection{Fundamentals}

Let $V$ be a vector space over $\cz$. An indefinite inner product on $V$ is a
map $(\cdot,\cdot): V \times V \to \cz$ which satisfies
\begin{gather*}
 (v,\lambda w_1 + \mu w_2)=\lambda (v,w_1) + \mu (v,w_2)\\
 \overline{(v_1,v_2)} = (v_2,v_1).
\end{gather*}
The indefinite inner product is said to be non-degenerate,
if
\begin{gather*}
 (v,w)=0, \quad \forall v \in V \;\Rightarrow w=0.
\end{gather*}
A non-degenerated indefinite inner product space $V$ is called decomposable
if it can be written as the direct sum of orthogonal subspaces $V^+$ and
$V^-$ such that the inner product is positive definite on $V^+$ and negative
definite on $V^-$. The inner product then defines a norm on these subspaces.
$V^+$ and $V^-$ are called intrinsically complete if they are complete in
these norms.
A non-degenerate indefinite inner product space which is decomposable such that
the subspaces $V^+$ and $V^-$ are intrinsically complete is called a {\bf Krein space}.
For every decomposition $V=V^+ \oplus V^-$ the operator $\J= \textrm{id} \oplus
-\textrm{id}$ defines a positive definite inner product (the $\J$-inner product) by $\langle \cdot ,
\cdot \rangle_\J:=(\cdot,\J \cdot)$. Such an operator $\J$ is called a fundamental symmetry.
It turns out that if $V$ is a Krein space each fundamental symmetry makes $V$
a Hilbert space. Furthermore two Hilbert space norms associated to different fundamental
symmetries are equivalent. The topology induced by these norms is called the
strong topology on $V$.
The theory of Krein spaces can be found in \cite{Bognar:1974}. For the sake
of completeness we will review in the following the main properties of linear
operators on Krein spaces.

\subsection{Operators on Krein spaces}

If $A$ is a linear operator on a Krein space $V$ we say that
$A$ is densely defined if the domain of definition $\mathcal{D}(A)$ of $A$
is strongly dense in $V$.
Let $A$ be a densely defined operator on a Krein space $V$. We may define the
Krein adjoint $A^+$ in the following way.
Let $\mathcal{D}(A^+)$ be the set of vectors $v$, such that there is a vector
$v^+$ with 
\begin{gather}
 (v, A w) = (v^+,w) \quad \forall w \in \mathcal{D}(A).
\end{gather}
We set $A^+ v := v^+$.
A densely defined operator is called Krein-selfadjoint if $A=A^+$.
An operator is called closed if its graph is closed in the strong topology,
i.e. if the operator is closed as an operator on the Hilbert space associated
to one (and hence to all) of the fundamental symmetries. If the closure of the
graph of an operator $A$ in the strong topology is an operator graph, then
$A$ is called closeable, the closure $\overline A$ is the operator associated
with the closure of the operator graph.
It turns out that a densely defined operator $A$ is closeable if and only if
$A^+$ is densely defined. The closure of $A$ is then given by $\overline A =
A^{++}:=(A^+)^+$. Clearly, a Krein-selfadjoint operator is always closed.
A densely defined operator is called essentially Krein-selfadjoint if it
is closable and its closure is Krein-selfadjoint.
Note that for any fundamental symmetry we have the equality $A^+=\J A^* \J$,
if the star denotes the adjoint in the Hilbert space defined by the $\J$-inner
product. Therefore, given a fundamental symmetry $\J$ and a Krein-selfadjoint
operator $A$, the operators $\J A$ and $A\J$ are selfadjoint as operators
in the Hilbert space induced by the $\J$-inner product.
The symmetric operators $\textrm{Re}(\frac{1}{2}(A + A^*))$ and
\mbox{$\textrm{Re}(\frac{\I}{2}(A - A^*))$} are called real and imaginary parts of
$A$. The sum of the squares of these operators is formally given by
$(A)_\J := \frac{1}{2}(A^* A + A A^*)$. It is natural to define the
$\J$-modulus of $A$ as its square root. For a special class of
fundamental symmetries this can be done straightforwardly.
We have
\begin{pro}
 Let $A$ be a Krein-selfadjoint operator on a Krein space $V$.
 Let $\J$ be a fundamental symmetry, such that $\dom(A) \cap \J \dom(A)$
 is dense in $V$. Let $\langle\cdot,\cdot \rangle$  be the scalar product
 associated with $\J$. Then the quadratic form 
 \begin{gather}
  q(\phi_1,\phi_2):=\frac{1}{2}(\langle A \phi_1 , A \phi_2
  \rangle + \langle A^* \phi_1 , A^* \phi_2\rangle)
 \end{gather}
 on $\dom(A) \cap \J \dom(A)$ is closed and the unique selfadjoint operator $(A)_\J$
 associated with this form commutes with $\J$. Therefore, it is
 Krein-selfadjoint. Moreover $\dom((A)_\J^{1/2})=\dom(A) \cap \J \dom(A)$.
\end{pro}
\begin{proof}
 Since both $A$ and $A^*$ are closed, the quadratic form is closed as well.
 We repeat the construction of the
 selfadjoint operator associated with this form (see \cite{RS:01}, Theorem
 VIII.15). Denote $W=\dom(A) \cap \J \dom(A)$.
 The pairing of the scalar product yields an inclusion of spaces $W \subset V \subset W^*$,
 where $W^*$ is the dual space of $W$. We define the operator $\hat B: W
 \to W^*$ by $[\hat B \phi](\psi):=q(\psi,\phi)+\langle \psi, \phi \rangle$.
 $\hat B$ is in isometric isomorphism. With $\dom(B):=\{ \psi \in W ; \hat B
 \psi \in V \}$ the operator $B:=\hat B \vert_{\dom(B)}: \dom(B) \to V$ is
 selfadjoint and $(A)_\J=B-1$. By construction $\J$ restricts on $W$ to a norm preserving
 isomorphism and its adjoint map $\hat \J^* : W^* \to W^*$ is the continuous
 extension of $\J$. From the definition of $\hat B$ we get immediately $\hat B
 \J = \hat \J^* \hat B$. Therefore, $\dom((A)_\J)=\dom(B)$ is invariant under
 the action of $\J$ and furthermore $(A)_\J$ and $\J$ commute.
 The form domain of $(A)_\J$ is $\dom((A)_\J^{1/2})$, and we conclude that
 $\dom((A)_\J^{\frac{1}{2}})=W$.
\end{proof}\noindent
We therefore define
\begin{definition}
 Let $A$ be a Krein-selfadjoint operator on a Krein space $V$ and suppose
 that $\J$ is a fundamental symmetry such that $\dom(A) \cap \J \dom(A)$
 is dense in $V$. Then the $\J$-modulus $[A]_\J$ of $A$ is the
 Krein-selfadjoint operator $(A)_\J^{1/2}$ constructed above.
\end{definition}

\subsection{Ideals of operators on Krein spaces}

Since all $\J$-inner products define equivalent norms, properties of
operators like
boundedness and compactness, which depend only on the topological structure
of the Hilbert space, carry over to Krein spaces without change.
The algebra of bounded operators in a Krein space $V$ will be denoted by
$\mathcal{B}(V)$. Each fundamental symmetry $\J$ defines a norm on
$\mathcal{B}(V)$ by $\Vert a \Vert_{\J}:= \sup_{v} \frac{\Vert a v
  \Vert_\J}{\Vert v \Vert_\J}$, where $\Vert v \Vert_\J^2=(v,\J v)$.
The norms on $\mathcal{B}(V)$ induced by different fundamental symmetries are
equivalent. We choose a fundamental symmetry $\J$ and view $V$ as a Hilbert
space with the $\J$-inner product.
Since $\mathcal{L}^{p+}$ are ideals in $\mathcal{B}(V)$,
we have $B^{-1} \mathcal{L}^{p+} B = \mathcal{L}^{p+}$ for any invertible operator in
$\mathcal{B}(V)$. Therefore, the definition of $\mathcal{L}^{p+}$
does not depend on the choice of scalar product and consequently it is
independent of the chosen fundamental symmetry.
The same argument applies to the Dixmier traces.
Let $\omega \in \Gamma_s(l^\infty)$ be fixed. Then for any $a \in
\mathcal{L}^{1+}$ and any invertible operator in $\mathcal{B}(V)$
we have $\Dix(B^{-1} a B) = \Dix(a)$. Therefore, the Dixmier trace does
not depend on the choice of fundamental symmetry.
We conclude that both $\mathcal{L}^{p+}$ and $\Dix$ make sense on Krein spaces
without referring to a particular fundamental symmetry.

\section{Clifford algebras and the Dirac operator}

\subsection{Clifford algebras and the spinor modules}
Let $q_{n,k}$ be the quadratic form $q(x)=-x_1^2-\ldots-x_k^2+x_{k+1}^2+\ldots+x_{n}^2$ on
$\rz^n$. The Clifford algebra $\Cl_{n,k}$ is the algebra generated by the
symbols $c(x)$ with $x \in \rz^n$ and the relations
\begin{gather}
 x \to c(x) \textrm{ is linear,}\\
 c(x)^2= q_{n,k}(x) 1.
\end{gather}
Let $\Clc_{n,k}$ be the complexification of $\Cl_{n,k}$
endowed with the antilinear involution ${}^+$ defined by
$c(v)^+=(-1)^k c(v)$.
For $n$ even the algebra $\Clc_{n,k}$ is isomorphic
to the matrix algebra $\Mat(2^\frac{n}{2})$, for $n$ odd
it is isomorphic to $\Mat(2^{[\frac{n}{2}]}) \oplus \Mat(2^{[\frac{n}{2}]})$.
Let $\sigma_1,\sigma_2,\sigma_3$ be the Pauli matrices and define
$$
 \tau(j) = \left \{ \begin{array}{cc} \mathrm{i}\; \textrm{for}\; j \leq k\\
     1 \;\textrm{for}\; j>k. \end{array} \right.
$$
For $n$ even we define the isomorphism $\Phi_{n,k}: \Clc_{n,k} \to
\Mat(2^\frac{n}{2})$ by
\begin{gather*}
 \Phi_{n,k}(c(x_{2j+1})):=\tau(2j+1).\underbrace{\sigma_3 \otimes \ldots \otimes \sigma_3}_{j-\textrm{times}}
 \otimes \sigma_1 \otimes 1 \otimes \ldots \otimes 1, \\
 \Phi_{n,k}(c(x_{2j})):=\tau(2j).\underbrace{\sigma_3 \otimes \ldots \otimes \sigma_3}_{(j-1)-\textrm{times}}
 \otimes \sigma_2 \otimes 1 \otimes \ldots \otimes 1.
\end{gather*}
Whereas for odd $n=2m+1$ we define $\Phi_{n,k}: \Clc_{n,k} \to
\Mat(2^{[\frac{n}{2}]}) \oplus \Mat(2^{[\frac{n}{2}]})$ by
\begin{gather*}
 \Phi_{n,k}(c(x_{j})):= \left \{ \begin{array}{cc}
 \Phi_{2m,k}(c(x_j)) \oplus \Phi_{2m,k}(c(x_j)) \; \textrm{for} \;
 1 \leq j\leq 2m, \\
 \tau(j) \{(\sigma_3 \otimes \ldots \otimes \sigma_3) \oplus  (- \sigma_3 \otimes
 \ldots \otimes \sigma_3)\} \; \textrm{for} \; j=2m+1. \\ \end{array} \right.
\end{gather*}
For even $n$ the isomorphism $\Phi_{n,k}$ gives an irreducible representation
of $\Clc_{n,k}$ on $\Delta_{n,k}:=\cz^{2^\frac{n}{2}}$, whereas for $n$ odd
we obtain an irreducible representation on
$\Delta_{n,k}:=\cz^{2^{[\frac{n}{2}]}}$ by restricting $\Phi_{n,k}$ to the
first component. The restrictions of these representations to the group
$\Spin(n,k) \subset \Cl_{n,k}$ are the well known spinor representations
on $\Delta_{n,k}$. In the following we write $\gamma(v)$ for
the image of $c(v)$ under this representation.
In case $n$ is even we define the grading operator $\chi:=\mathrm{i}^{\frac{n(n-1)}{2}+k}\gamma(x_1) \cdots
\gamma(x_n)$.
We have
\begin{gather}
\chi^2=1, \\
\chi \gamma(v) + \gamma(v) \chi = 0.
\end{gather}
There is no analogue to
this operator in the odd dimensional case.
There is a natural non-degenerate indefinite inner product on the modules $\Delta_{n,k}$
given by
\begin{gather}
 (u,v)=\I^{\frac{k(k+1)}{2}} \langle \gamma(x_1) \cdots \gamma(x_k) u, v \rangle_{\cz^{2^{[\frac{n}{2}]}}}.
\end{gather}
This indefinite inner product is invariant
under the action of the group $\Spin(n,k)^+$ which is the
double covering group of $\SO(n,k)^+$.
Furthermore the Krein-adjoint $\Phi_{n,k}(x)^+$ of $\Phi_{n,k}(x)$
is given by $\Phi_{n,k}(x^+)$. If $n$ is even, one gets for the grading
operator $\chi^+=(-1)^k \chi$. Up to a factor this inner product is uniquely
determined by these properties.

\subsection{Fundamental symmetries of the spinor modules}
Let now $n$ and $k$ be fixed and denote by $g$ the unique bilinear form on $\rz^n$
such that $g(v,v)=q_{n,k}(v)$. A spacelike reflection is linear map $r: \rz^n
\to \rz^n$ with $r^2=1$, $g(r u, r v) = g(u,v)$ for all $u,v \in \rz^n$
such that $g(\cdot, r \cdot)$ is a positive definite inner product.
Each such reflection determines a splitting $\rz^n=\rz^{k} \oplus \rz^{n-k}$
into $g$-orthogonal eigenspaces of $r$ for eigenvalue $-1$ and $+1$.
Clearly, $g$ is negative definite on the first and positive definite on the
second summand. Conversely for each splitting of $\rz^n$ into a direct
sum of $g$-orthogonal subspaces such that $g$ is negative or positive
definite on the summands determines a spacelike reflection.

To each such spacelike reflection we can associate a fundamental symmetry
of the Krein space $\Delta_{n,k}$.
We choose an oriented orthonormal basis $(e_1,\ldots,e_k)$
in the eigen\-space for eigenvalue $-1$. Then the operator
$\J_r:=\I^{\frac{k(k+1)}{2}} \gamma(e_1) \cdots \gamma(e_k)$
is a fundamental symmetry of $\Delta_{n,k}$ and
we have $\J_r \gamma(v) \J_r = (-1)^{k} \gamma(r v)$.
In general, not all the fundamental symmetries are of this form.
The following criterion will turn out to be useful.
\begin{pro}\label{scalar}
 Let $\J$ be a fundamental symmetry of the Krein space $\Delta_{n,k}$
 such that for each $v \in \rz^n \subset \Cl^c_{n,k}$ the matrix
 $$
  (\J \gamma(v))^2 + (\gamma(v) \J)^2
 $$
 is proportional to the identity. If $n$ is even assume furthermore
 that $\J$ commutes or anticommutes with the grading operator $\chi$.
 Then there is a spacelike reflection $r$ such that $\J=\J_r$.
\end{pro}
\begin{proof}
 By assumption $h(u,v)=(-1)^k \{ \J \gamma(u) \J , \gamma(v) \}=\{
 \gamma(u)^*,\gamma(v) \}$
 is a real valued bilinear form on $\rz^n$, where $\{ \cdot, \cdot\}$
 denotes the anti-commutator and the ${}^*$ is the adjoint in the $\J$-scalar
 product. Clearly, $h(v,v) \geq 0$ for all $v \in \rz^n$.
 Therefore, there exists a
 matrix $a \in \textrm{End}(\rz^n)$ such that
 \begin{gather}
  \{ \gamma(u)^*, \gamma(v) \} =  \{ \gamma(a u), \gamma(v) \}
 \end{gather}
 for all $u,v \in \rz^n$.
 As a consequence $\delta(u)=\gamma(u)^*-\gamma(a u)$ anti-commutes
 with all elements $\gamma(v)$. In the odd dimensional case there
 is no such matrix other than $0$ and in the even dimensional case
 $\delta$ must be a multiple of the grading operator.
 Therefore, $\gamma^*(v)=(-1)^k \J \gamma(v) \J$ can be written as a sum
 $\gamma(a v) + \lambda(v) \chi$, where $\lambda$ is a linear
 form on $\rz^n$.  From $\gamma(v)^{**}=\gamma(v)$ and $\J \chi \J = \pm \chi$ we get
 $a^2=1$. For eigenvectors $a v = \pm v$ of $a$ one gets from the equation
 $(\gamma(v)^*)^2=\gamma(v)^2$ that $\lambda(v)^2=0$. Hence, $\lambda=0$.
 We showed that for $n$ even or odd we always have
 $\J \gamma(v) \J = (-1)^{k} \gamma(a v)$, for a reflection $a$.
 The bilinear form $h$ is $h(u,v)=(u,a v)$
 and since it is positive semi-definite and $a$ has trivial kernel,
 it is positive definite. Therefore, $a$ is a spacelike reflection
 and consequently $\J_a \gamma(v) \J_a = (-1)^{k}\gamma(a v)$.
 It remains to show that $\J_a=\J$.
 From the above relation one gets $\J \J_a \gamma(v) \J_a \J = \gamma(v)$,
 and therefore, $\J \J_a$ commutes with all $\gamma(v)$ and has to be a multiple
 of the identity. Hence, $\J= z \J_a$ for some complex number $z$.
 From $\J^2=\J_a^2=1$, $\J_a^+=\J_a$ and $\J^+=\J$ we get $z=\pm 1$.
 Since the fundamental symmetries both give rise to positive definite
 scalar products on $\Delta_{n,k}$, we conclude that $z=1$ and $\J=\J_a$.
\end{proof}

\subsection{Semi-Riemannian geometry and the Dirac operator}

Let $M$ be a smooth $n$-dimensional manifold.
A semi-Riemannian metric $g$ on $M$ is a smooth section in the bundle $T^*M \otimes T^*M$,
such that for all $x \in M$ the bilinear form
$g_x$ on $T_x^*M \times T_x^*M$ is non-degenerate. If $g_x(v,v)=q_{n,k}(v)$
for a special choice of basis we say that $g_x$ has signature $(n,k)$.
If $g_x$ has signature $(n,k)$ for all $x \in M$ the metric is called
semi-Riemannian. If the signature is $(n,0)$ then the metric is called
Riemannian, in case the signature is $(n,1)$ the metric is called Lorentzian.
A vector field $\xi$ is called timelike (spacelike, lightlike)
if $g(\xi,\xi) < 0,\;(>0, =0)$. The metric can be used to identify
$T^*M$ and $TM$ and therefore, $g$ can be regarded as a section in $TM \otimes
TM$ inducing a scalar product on $T^*_xM$. See \cite{Neill:1983} or
\cite{Baum} for elementary properties of semi-Riemannian manifolds.

If $(M,g)$ is a semi-Riemannian metric of signature $(n,k)$, then the
tangent-bundle $TM$ can be split into an orthogonal direct sum
$TM=F_1^k \oplus F_2^{n-k}$, where $g$ is negative definite on $F_1^k$
and positive definite on $F_2^{n-k}$. For such a splitting we can define
a map $r: TM \to TM$ by $r (x,k_1 \oplus k_2) := (x,-k_1 \oplus k_2)$.
Then the metric $g^r$ defined by $g^r(a,b):=g(a,r b)$ is positive definite.
Conversely suppose there is an endomorphism of vector bundles $r: TM \to TM$
with $g(r \cdot,r \cdot)=g$, $r^2 = \textrm{id}$ and such that $g^r:=g(\cdot,r \cdot)$
is positive definite. Then there is a splitting such that
$r (x,k_1 \oplus k_2) = (x,-k_1 \oplus k_2)$. We call such maps
spacelike reflections. Obviously, $r: TM \to TM$ is a spacelike reflection, if the restrictions
of $r$ to the fibres $T_x M$ are spacelike reflections in the sense
of the last section. In the following we call $g^r$ the Riemannian metric associated with $r$.

In case the bundle $TM,\;(F_1^{k},F_2^{n-k})$ is orientable the manifold
is called orientable (time-orientable, space-orientable).
Assume we are given an orientable, time-orientable semi-Riemannian manifold
$(M,g)$ of signature $(n,k)$. Then the bundle of oriented orthonormal frames
is an $\SO(n,k)^+$-principal bundle.

We saw that the metric information of a Riemannian manifold can be encoded
in a spectral triple, where $D$ was any Dirac type operator on some
hermitian vector bundle $E$.
In the case of semi-Riemannian manifolds there arises a major problem.
Namely that Dirac type operators are not selfadjoint any more.
We will see however that there exists a Krein space structure
on the space of sections of $E$ such that there are Krein-selfadjoint
Dirac type operators.
Assume now that $M$ is an orientable time-orientable semi-Riemannian manifold.
Let $E$ be a vector bundle over $M$ and assume that $D$ is of Dirac type.
This means that $D$ is a first order differential operator and
the principal symbol $\sigma$ of $D$ satisfies the
relation
\begin{gather}
 \sigma(\xi)^2 = g(\xi,\xi) \; \textrm{id}_{E_p} \quad \forall \xi \in T_p^*M, p \in M.
\end{gather}
Therefore, $\gamma:=\sigma$ satisfies the Clifford relations, which
makes $E$ a module for the Clifford algebra bundle.
Let $r : TM \to TM$ be a spacelike reflection and identify $TM$ with $T^*M$
using the metric. Let $T^*M=F_1^k \oplus F_2^{n-k}$ be the splitting such that
$r (x,k_1 \oplus k_2) = (x,-k_1 \oplus k_2)$. Then there is a hermitian
structure $\langle\cdot,\cdot\rangle$ on $E$ such that $\sigma_x(\xi)$ is
anti-symmetric if $\xi \in F_1^k$ and symmetric if \mbox{$\xi \in F_2^{n-k}$}. 
Let $e_1,\ldots,e_k$ be a local oriented orthonormal frame for $F_1^k$ and define
$\J := \I^{\frac{k(k+1)}{2}} \gamma(e_1) \cdots \gamma(e_k)$. $\J$ is independent of the choice
of frames and the indefinite inner product $(\cdot,\cdot)_x := \langle
\cdot,\J(x)\cdot\rangle_x$ on $E_x$ is non-degenerate. It makes $E$ an
non-degenerate indefinite inner product bundle.
Moreover, $\I^{k}\sigma_x$ is symmetric with respect to this indefinite inner product.
The space of square integrable sections of $E$ is a Krein space endowed with
the indefinite inner product structure
\begin{gather}
  (f,g):=\int_M (f_x,g_x)_x \sqrt{\vert g \vert} dx.
\end{gather}
To each spacelike reflection $r'$ we can associate a fundamental symmetry $\J_{r'}$ of
this Krein space by
$\J_{r'}:= \I^{\frac{k(k+1)}{2}}\gamma(e_1) \cdots \gamma(e_k)$, where
$e_1,\ldots,e_k$ is a local oriented orthonormal frame for $F_1^k$.
We conclude that for a time-orientable orientable
semi-Riemannian manifold there exists a Dirac type operator $D$ on some non-degenerate
indefinite inner product vector bundle $E$ such that $\I^{k} D$ is symmetric with
respect to this inner product.
The following theorem was proved for Dirac operators on spin manifolds in
\cite{Baum}. For the sake of completeness and since the original proof is in
german, we give a proof here.

\begin{theorem}[Baum, 1981]
 Let $E$ be a non-degenerate indefinite inner product vector bundle over an orientable time-orientable
 semi-Riemannian manifold $M^{n,k}$.
 Let $D: \Gamma_0(E) \to \Gamma_0(E)$ be a symmetric differential operator such that
 $\I^{k}D$ is of Dirac type.
 If there exists a spacelike reflection $r$ such that the Riemannian
 metric associated with this reflection is complete, then
 $D$ is essentially Krein-selfadjoint.
 In particular, if $M$ is compact then $D$ is always essentially Krein-selfadjoint.
\end{theorem}

\begin{proof}
 Let $\J$ be the fundamental symmetry associated with the splitting
 and let $L^2(E)$ be the Hilbert space of sections which are square integrable
 with respect to the positive definite inner product induced by $\J$.
 We denote this scalar product in the following by $\langle \cdot, \cdot
 \rangle$. It is clearly sufficient to show that $P = \J D$ is essentially
 selfadjoint in $L^2(E)$. Note that $P$ is a first order differential
 operator which is symmetric in $L^2(E)$. Therefore, it is closeable.
 The proof consists of two steps. Let $\dom_0(P^*)$ be the intersection
 of $\dom(P^*)$ with the space of compactly supported square integrable
 section. We first show that $\dom_0(P^*) \subset \dom(\overline P)$.
 In the second step we show that $\dom_0(P^*)$ is dense in the Hilbert
 space $\dom(P^*)$ endowed with the scalar product
 $\langle x,y \rangle_{P^*}:=\langle x,y \rangle + \langle P^* x , P^* y \rangle $.
 The combination of these results shows that
 $\dom(\overline P)$ is dense in the Hilbert space $\dom(P^*)$, and
 therefore, $P$ is essentially selfadjoint.\\
 {\bf First step:} Note that since $P$ is symmetric, the operator $P^*$ is a closed
 extension of $P$, and furthermore the adjoint operator $P': \Di(E) \to
 \Di(E)$ is the continuous extension of $P$ and of $P^*$ to the space of distributions.
 Assume that $f$ is in $\dom_0(P^*)$. Then both $f$ and $g=P^* f$ have compact
 support. Clearly, $f$ is a weak solution to the equation $P f = g$, hence,
 it is also a strong solution (see e.g. \cite{Taylor:1981}, Prop. 7.4), i.e.
 there is a sequence $f_n$ converging to $f$ in the $L^2$-sense such that
 $P f_n$ converges to $g$ also in the $L^2$-sense.
 Therefore, $f$ is in $\dom(\overline P)$.\\
 {\bf Second step:}
 Assume that $f \in \dom(P^*)$. We will construct a sequence $f_n$ in
 $\dom_0(P^*)$ such that $f_n \to f$ and $P^* f_n \to P^* f$ in the $L^2$-sense.
 Fix an $x_0 \in M$ and let $\textrm{dist}(x)$ be a regularized distance function
 from $x_0$ in the complete Riemannian metric associated with the splitting.
 Choose a function $\chi \in \co(\rz)$ with $0 \leq \chi \leq 1\;$,
 $\chi(t)=0$ for $t \geq 2$, $\chi(t)=1$ for $t \leq 1$, and $\vert
 \chi'\vert\leq 2$.
 We set $\chi_n(x):=\chi(\frac{1}{n} \textrm{dist}(x))$. By completeness of the manifold, all
 $\chi_n$ are compactly supported. We define the sequence $f_n:=\chi_n f$ and
 clearly, $f_n \in \dom_0(P^*)$. Denoting by $\sigma$ the principal symbol of
 $D$, we have 
 $P' f_n = -\I \J \sigma(d \chi_n) f + \chi_n P' f$. Clearly, $\chi_n P' f \to P'f$
 in the $L^2$-sense. For the first summand we have the estimate
 \begin{gather}
  \Vert \J \sigma(d \chi_n) f \Vert^2 \leq \int_{B_{2n}-B_n} \frac{4}{n^2}
  \Vert f \Vert^2,
 \end{gather}
 where $B_r$ denotes the metric ball with radius $r$ centred at $x_0$.
 Since the right hand side vanishes in the limit $n \to \infty$,
 we conclude
 that $f_n \to f$ and $P' f_n \to P'f$ in the $L^2$-sense. Therefore,
 $\dom_0(P^*)$ is dense in the Hilbert space $\dom(P^*)$.
\end{proof}

\begin{exa}
 A spin structure on a time-oriented oriented semi-Riemannian manifold  
 $M^{n,k}$ is an $\Spin(n,k)^+$-principal bundle $P$ over $M^{n,k}$ together with a smooth
 covering $\eta$ from $P$ onto the bundle $Q$ of oriented orthonormal frames,
 such that the following diagram is commutative.
 \begin{equation}
  \begin{CD}
    P \times \Spin(n,k)^+ @>>> P @>>> M^{n,k} \\
    @VV \eta \times \lambda  V  @VV \eta V   @| \\
    Q \times \SO(n,k)^+   @>>> Q @>>> M^{n,k}
  \end{CD}
 \end{equation}
 Here $\lambda$ denotes the covering map $\Spin(n,k)^+ \to \SO(n,k)^+$.
 The spinor bundle $S$ associated with a Spin structure is the associated bundle
 $P \times_\pi \Delta_{n,k}$, where $\pi$ denotes the representation of $\Spin(n,k)^+$ 
 on $\Delta_{n,k}$.
 Let $\nabla: \Gamma(S) \to \Gamma(S) \otimes \Lambda^1$ be the Levi-Civita
 connection on the spinor bundle. The Dirac operator $\Dirac$ is defined by 
 $-\I\gamma \circ \nabla$, where $\gamma$ denotes the action of covector fields
 on sections of the spinor bundle by Clifford multiplication.
 $\Dirac$ is clearly of Dirac type and it was shown in \cite{Baum} (see also
 \cite{Baum2}) that the space of square
 integrable sections of $S$ is a Krein space such that $\I^{k} \Dirac$ is symmetric. 
\end{exa}

\section{Semi-Riemannian spectral triples}

\begin{definition}
 A semi-Riemannian spectral triple is a tuple $(\alg,D,\Hilbert)$, where
 $\alg$ is a $*$-algebra of bounded operators on a Krein space $\Hilbert$
 such that $a^*=a^+$, and
 $D$ is a Krein-selfadjoint operator on $\Hilbert$, such that
 the commutators $[D,a]$ are bounded for all $a \in \alg$.
 A semi-Riemannian spectral triple is called even if there is a distinguished operator $\chi$,
 anticommuting with $D$ and commuting with $\alg$ and with $\chi^2=1$ and
 $\chi^+= \pm \chi$. If such an operator does not exist we say the spectral
 triple is odd and set by definition $\chi=1$. We call $\chi$ the grading operator.
 \end{definition}

For a semi-Riemannian spectral triple we can repeat the construction
of differential forms almost unchanged. Denote again the
universal differential envelope of $\alg$ by $(\Omega\alg,d)$.
Clearly, the map
\begin{gather*}
 \pi : \Omega \alg \to \mathcal{B}(\Hilbert),\\
 \pi((a_0,a_1,\ldots,a_n)):= a_0 [D,a_1] \cdots [D,a_n], \quad a_j \in \alg
\end{gather*}
is a representation of $\Omega\alg$ on $\Hilbert$ such that
$\pi(a^*)=\pi(a)^+$ for all $a \in \Omega\alg$, where ${}^+$ denotes the Krein adjoint.
We define the graded two sided ideal $j_0:=\oplus_n j_0^n$ by
\begin{gather}
 j_0^n := \{ \omega \in \Omega^n \alg ; \pi(\omega)=0 \},
\end{gather}
and as in the case of spectral triples $j= j_0 + d j_0$
is a graded two-sided differential ideal.
We define $\Omega_D\alg:=\Omega\alg / j$. Clearly, $\Omega_D^n \alg \simeq
\pi(\Omega^n \alg) / \pi(d j_0 \cap \Omega^n \alg)$.
\begin{exa}
 Suppose that $M^{n,k}$ is a compact time-orientable orientable
 semi-Riemannian spin manifold with spinor bundle $E$.
 Let $\Hilbert$ be the Krein
 space of square integrable sections of $E$ and let $D=\I^{k}\Dirac$,
 where $\Dirac$ is the Dirac
 operator. Then the triple $(C^\infty(M),D,\Hilbert)$ is a semi-Riemannian spectral
 triple, and in the same way as this is done for the Riemannian case one shows
 that as a graded differential algebra $\Omega_D C^\infty(M)$
 is canonically isomorphic to the algebra of differential forms on $M$.
 We call this triple the canonical triple associated with $M$.
\end{exa}

For Riemannian spin manifolds the differential structure is encoded in the
Dirac operator. For example the space of smooth sections of the spinor bundle
coincides with the space $\bigcap_{n} \dom(D^n)$. This is essentially due
to the ellipticity of the Dirac operator. In the semi-Riemannian case the
Dirac operator is not elliptic any more and sections in $\bigcap_{n}
\dom(D^n)$ may be singular in the lightlike directions. We will circumvent
this problem by introducing the notion of a smooth
semi-Riemannian spectral triple.
Let $(\alg,D,\Hilbert)$ be a semi-Riemannian spectral triple and suppose
there is a fundamental symmetry, such that $\dom(D) \cap \J \dom(D)$
is dense in $\Hilbert$. Then the operator $\Delta_\J:=([D]_\J^2+1)^{1/2}$
is a selfadjoint operator on the Hilbert space $\Hilbert$ with scalar
product $(\cdot,\J\cdot)$. Let $\Hilbert_\J^s$ be the closure
of $\bigcap_{n} \dom(\Delta_\J^n)$ in the norm $\Vert \psi \Vert_s:=\Vert
\Delta_\J^{s} \psi \Vert$. We have for $s>0$ the equality
$\Hilbert^s_\J=\dom(\Delta_\J^s)$, and the indefinite inner product
on $\Hilbert$ can be used to identify $\Hilbert^s_\J$ with the topological
dual of $\Hilbert^{-s}_\J$. We define $\Hilbert_\J^{\infty}:=\bigcap_{s}
\Hilbert^s_\J$ and $\Hilbert_\J^{-\infty}:=\bigcup_{s}
\Hilbert^s_\J$. A map $a: \Hilbert^{-\infty} \to \Hilbert^{-\infty}$
is said to be in $\op_\J^r$ if it continuously maps $\Hilbert^s$
to $\Hilbert^{s-r}$. Clearly, $\Delta_\J \in \op^1_\J$ and $\J \in \op^0_\J$.
We introduce an equivalence relation on the set of fundamental symmetries $\J$
such that $\dom(D) \cap \J \dom(D)$ is dense in $\Hilbert$ in the following
way. We say that $\J_1 \sim \J_2$ if $\Hilbert^s_{\J_1}=\Hilbert^s_{\J_2}$
as topological vector spaces.
If we are dealing with a distinguished equivalence class, we will leave away
the index $\J$ and write e.g $\op^r$ for $\op^r_\J$ and $\Hilbert^s$
for $\Hilbert^s_\J$, since these objects clearly depend only on the
equivalence class of $\J$. The spaces $\op^r$ have been introduced in \cite{Connes:1995hg}
in the context of spectral triples.

\begin{definition}
 A smooth semi-Riemannian spectral triple is a semi-Riemannian spectral triple
 $(\alg,D,\Hilbert)$ together with a distinguished non-empty equivalence class
 of fundamental symmetries $[\J]$, such that $D \in \op^1$. We say a
 fundamental symmetry is smooth if $\J \in [\J]$.
\end{definition}

\begin{exa}\label{exasmooth}
 Let $M^{n,k}$ be a compact orientable time-orientable semi-Riemannian spin manifold
 and let $(\alg,D,\Hilbert)$ be its canonical semi-Riemannian spectral
 triple. Let $E$ be the spinor bundle. For each spacelike reflection $r$ we
 constructed in the previous section a fundamental symmetry $\J_r$ of the
 Krein space $\Hilbert$.
 The fundamental symmetries of the form $\J_r$
 belong to one and the same equivalence class
 and therefore define a smooth semi-Riemannian spectral triple.
 This can most easily be seen using the calculus of pseudo\-differential
 operators. Note that $\Delta_\J$ is an elliptic classical
 pseudo\-differential operator of order $1$. Therefore, $\Hilbert^s_\J$ coincides
 with the
 Sobolev space $H_s(M,E)$ of sections of $E$ and $\Hilbert^\infty$
 coincides with the space of smooth sections $\Gamma(E)$.
 In the following we will think of the canonical triple associated with
 $M$ as a smooth semi-Riemannian spectral triple with the above smooth structure.
\end{exa}

Suppose that $(\alg,D,\Hilbert)$ is a smooth semi-Riemannian spectral triple.
Let $\J_1$ and $\J_2$ be smooth fundamental symmetries. Then $\Delta_{\J_1}^{-1}$
is in $\op^{-1}$ and therefore, $\Delta_{\J_2} \Delta_{\J_1}^{-1}$ is bounded.
As a consequence $\Delta_{\J_1}^{-1}$ is in $\mathcal{L}^{p+}$ if and only if
$\Delta_{\J_1}^{-1}$ is in $\mathcal{L}^{p+}$.
\begin{definition}
 We say a smooth semi-Riemannian spectral triple $(\alg,D,\Hilbert)$ is $p^+$-summable
 if for one (and hence for all) smooth fundamental symmetries $\J$ the operator
 $\Delta_\J^{-1}$ is in $\mathcal{L}^{p+}$.
\end{definition}

For the canonical triple associated with a semi-Riemannian spin manifold
we have a distinguished set of fundamental symmetries, namely those
which are of the form $\J=\J_r$ for some spacelike reflection $r$.
We may ask now if there is an analogue of this set in the general case.

\begin{definition}
 Let $(\alg,\Hilbert,D)$ be a smooth semi-Riemannian spectral triple.
 We say a fundamental symmetry $\J$ is admissible if
 \begin{enumerate}
  \item $\J$ is smooth,
  \item $\J \chi \J = \chi^+$,
  \item $\J$ commutes with all elements of $\alg$,
  \item $\J \pi(\Omega^p\alg) \J = \pi (\Omega^p \alg)$.
  \item $\J \pi(d j_0 \cap \Omega^p \alg) \J = \pi(d j_0 \cap \Omega^p \alg)$
    if $p \geq 2$.
 \end{enumerate}
\end{definition}
If $\J$ is admissible and $^*$ denotes the adjoint in the $\J$-inner product,
then the above conditions imply that $a^*=a^+$ for all $a \in \alg$, $\chi^*=\chi$ and that
$^*$ leaves the spaces $\Omega^p_D \alg$ invariant.
The following theorem shows that in the case of a canonical triple
associated with a spin  manifold
the set of admissible fundamental symmetries
is canonically isomorphic to the set of spacelike reflections.

\begin{theorem}\label{sstr}
 Suppose that $M^{n,k}$ is an orientable time-orientable compact semi-Riemannian
 spin manifold and let $(\alg,D,\Hilbert)$ be its canonical smooth
 triple with grading operator $\chi$.
 Then the set of admissible fundamental symmetries coincides with the set
  \begin{gather*}
  \{ \J_{r}\; ; r \textrm{ is a spacelike reflection} \}.
 \end{gather*}
\end{theorem}

\begin{proof}
 We first show that $\J_r$ is admissible.
 Clearly, $\J_r$ commutes with all elements of $\alg$ and
 $\J \chi \J = \chi^+$. Moreover $\J_r$ is smooth by construction (see
 Example \ref{exasmooth}). We need to show that the sets $\pi(\Omega^p \alg)$
 and $\pi(d j_0 \cap \Omega^p
 \alg)$ are invariant under conjugation by $\J_r$. We denote by $\gamma$
 the principal symbol of $D$. Since $[D,f]=-(\I)^{k+1}\gamma(df)$, the space
 $\pi(\Omega^p \alg)$ is the set of operators of the form
 $\sum_j f^j \gamma(v_1^j) \cdots \gamma(v_p^j)$, where $f^j \in C^\infty(M)$ and
 $v_1^j,\ldots,v_p^j \in \Gamma(T^*M)$. Since $\J_r \gamma(v) \J_r = (-1)^{k}
 \gamma(rv)$, this space is invariant under
 conjugation by $\J_r$. The proof of
 Proposition 7.2.2 in \cite{Landi:1997}
 shows that $\pi(d j_0 \cap \Omega^p \alg)$
 coincides with the set of operators of the form $\sum_j f^j \gamma(v_1^j) \cdots
 \gamma(v_{p-2}^j)$, where $f^j \in \alg$, $v_1^j,\ldots, v_k^j \in T^*M$.
 This set is also invariant under
 conjugation by $\J_r$ and we conclude that $\J_r$ is admissible.
 Suppose now we have another admissible fundamental symmetry $\J$ in $[\J_r]$.
 Since $\J \in \op^0$, it acts
 continuously on $\Gamma(E)$ and since $\J$ commutes with $\alg$, it leaves the fibres invariant.
 It follows that $\J$ is a smooth endomorphism of the spinor bundle.
 For a point $x \in M$ we denote by $\J(x)$ the restriction of $\J$ to the fibre
 at $x$. Since $\J$ is admissible, $\J \cdot \J$ must leave the space of one-forms
 invariant. This implies that for all $v \in T^*_x M$ the matrix
 $\J(x) \gamma(v) \J(x)$ is again of the form $\gamma(u)$ for some $u \in T^*_x
 M$. 
 By Prop. \ref{scalar} there exists a spacelike reflection on the fibre at $x$
 inducing $\J(x)$. Therefore, there is spacelike reflection $r$ such that $\J=\J_{r}$.
\end{proof}

\begin{theorem} \label{th2}
 Let $M^{n,k}$ be a compact orientable time-orientable semi-Riemannian
 spin manifold and let $(\alg,D,\Hilbert)$ be the canonical smooth
 semi-Riemannian spectral triple associated with $M$. Then
 $(\alg,D,\Hilbert)$ is $n^+$-summable and
 for each $f \in C^\infty(M)$ and each admissible fundamental symmetry $\J$ we have
 \begin{gather}\label{integ}
  \int_M f = c(n) \Dix (f \Delta_\J^{-n}),
 \end{gather}
 where integration is taken with
 respect to the semi-Riemannian volume form $\sqrt{\vert g \vert}$
 and $c(n)=2^{n-[n/2]-1}\pi^{n/2}n \Gamma(n/2)$.
 Moreover with the same $f$ and $\J$
 \begin{gather}\label{index}
  \Dix (f D^2 \Delta_\J^{-n-2})= (-1)^{k}\frac{n-2k}{n} \Dix (f \Delta_\J^{-n}).
 \end{gather}
\end{theorem}
\begin{proof}
 Let $g^r$ be the Riemannian metric
 associated with a spacelike reflection. By construction the
 metric volume form of $g^r$ coincides with the metric volume form of the
 semi-Riemannian metric. Now the principal symbol $\sigma_1$ of $\Delta_{\J_r}$
 is given by $\sigma_1(k)=\sqrt{g^r(k,k)}$ for covectors $k \in T^*M$.
 Connes´ trace formula gives equation \ref{integ}.
 What is left is to show that equation \ref{index} holds.
 The operator $D^2 \Delta_\J^{-n-2}$ is a classical pseudo\-differential
 operator of order $-n$ and its principal symbol $\sigma_2$ is given by
 $\sigma_2(k)=(-1)^{k}g(k,k) g^r(k,k)^{-n/2-1}$. Therefore, the principal symbol
 of $f D^2 \Delta_\J^{-n-2}$ is $f \sigma_2$. In order to calculate the
 relevant Dixmier trace we have to integrate this symbol over the
 cosphere bundle in some Riemannian metric. The result will be independent
 of the chosen Riemannian metric. In case $\J=\J_r$ we use $g^r$ to integrate.
 On the cosphere bundle $\sigma_2$ restricts to $g(k,k)$. Therefore,
 \begin{gather}
  \Dix (f D^2 \Delta_\J^{-n-2})=\frac{1}{c(n)} (-1)^{k}\mathrm{Vol}(S^{n-1})^{-1} \int_{S^*M} f \cdot g \; .
 \end{gather}
 For local integration we can choose an oriented orthonormal frame $k_1,\ldots,k_n$
 such that $g(k_i,k_i)=-1$ for $i=1 \ldots k$ and $g(k_i,k_i)=1$
 for $i=k+1 \ldots n$. This shows that
 \begin{gather}\nonumber
  \int_{S^*M} f \cdot g = (-1)^{k}\left( \int_M f \right) \cdot \int_{S^{n-1}} (-\xi_1^2-\ldots-\xi_k^2+
  \xi_{k+1}^2+\ldots+\xi_n) \\=  (-1)^{k}\mathrm{Vol}(S^{n-1})\frac{n-2k}{n} \int_M f \;,
 \end{gather}
 which concludes the proof.
\end{proof}

Equation \ref{index} shows that one can indeed recover the signature
from the spectral data and that the notion of integration
is independent of the chosen admissible fundamental symmetry.

The conditions for a fundamental symmetry to be admissible
are in a sense minimal and it is not clear at this point that
one does not need further conditions in order to get a sensible
noncommutative geometry. For example one may require in addition that
the set $\alg \cup [D,\alg]$ is contained in the domain of smoothness of the
derivation $\delta_\J(\cdot)=[\Delta_\J,\cdot]$. This is clearly true
for admissible fundamental symmetries in the case of a canonical
spectral triple associated with a manifold. In the general case however we can not
expect this to hold.
We think it is also worth noting that for the classical
situation there exist a number of equivalent definitions of admissibility.
For example one has

\begin{pro}
 Let $(\alg,D,\Hilbert)$ be as in Theorem \ref{th2}. Assume that $\J$ is a
 smooth fundamental symmetry that
 commutes with all elements of $\alg$ and $\J \chi \J = \chi^+$.
 Then $\alg \cup [D,\alg]$ is contained in
 the domain of the derivation $\delta_\J$ if and only if $\J$ is admissible.
\end{pro}
\begin{proof}
 By assumption $\J$ is a smooth endomorphism of the spinor bundle. Let $\J(x)$
 be the restriction to the fibre at $x$. Denote by $\sigma$ the principal
 symbol of $D$.
 The principal symbol $A$
 of the second order pseudo\-differential operator $\Delta_\J^2$ is given by
 $A_x(v)=\frac{1}{2}(\J \sigma_x(v) \J \sigma_x(v) + \sigma_x(v) \J \sigma_x(v) \J)$
 for $v \in T^*_xM$.
 The principal symbol of $\Delta_\J$ is $A^{1/2}$. Assume now that $[\Delta_\J,a]$
 is bounded for all $a$ in $\alg \cup [D,\alg]$. Then the principal symbol of
 the first order operator $[\Delta_\J,a]$ must vanish. This implies that $A^{1/2}_x(v)$
 commutes with all $\sigma_x(u);\; u \in T^*_X M$. Since the Clifford
 action is irreducible, $A_x(v)$ 
 is a multiple of the identity and by Prop. \ref{scalar} we have $\J=\J_r$
 for some spacelike reflection $r$.
\end{proof}

\section{The noncommutative tori}

\begin{definition}
 Let $\theta$ be a pre-symplectic form on $\rz^n$. We denote by $A_\theta$
 the unital $C^*$-algebra generated by symbols $u(y), \; y \in \zz^n$ and relations
 \begin{gather}
  u(y)^*=u(y)^{-1} \\
  u(y_1) u(y_2) = e^{\I \pi \theta(y_1,y_2)} u(y_1+y_2).
 \end{gather}
 Let $\mathcal{S}(\zz^n)$ be the Schwarz space over $\zz^n$, i.e. the space
 of functions on  $\zz^n$ with
 \begin{gather}
  \sup_{y \in \zz^n} (1+\vert y \vert^2)^p \vert a(y) \vert^2 < \infty \quad \forall p
  \in \nz.
 \end{gather}
 The rotation algebra $\alg_\theta$ is defined by
 \begin{gather}
  \alg_\theta := \left\{ a = \sum_{y \in \zz^n} a(y) u(y)\; ; \; a \in
  \mathcal{S}(\zz^n) \right\}.
 \end{gather}
\end{definition}

It is well known that the linear functional $\tau: \alg_\theta \to \cz$
defined by
\begin{gather}
  \tau \left( \sum_r a(y) u(y) \right) := a(0),
 \end{gather}
is a faithful tracial state over $\alg_\theta$.
In particular we have $\tau(a^* a)=\sum_{y \in \zz^n} \vert a(y) \vert^2$.
Note that $A_\theta$ is generated by the elements $u_k:=u(e_k)$,
where $e_k$ are the basis elements in $\zz^n$.
They satisfy the relations
\begin{gather}
 u_k^*=u_k^{-1}\\
 u_k u_i = e^{2 \pi \I \theta_{ik} } u_i u_k
\end{gather}

\begin{definition}
 The basic derivations $\delta_1,\ldots,\delta_n$ on $\alg_\theta$
 are defined by
 \begin{gather}
  \delta_j \left( \sum_{y \in \zz^n} a(y) u(y)\right):= 2 \pi \I \sum_{y \in \zz^n} y_j
  a(y) u(y).
 \end{gather}
 One checks easily that these are indeed derivations.
\end{definition}

Let $\Hilbert_\tau$ be the GNS-Hilbert space of the state $\tau$.
Since $\tau$ is faithful, $\Hilbert_\tau$ coincides with the
closure of $\alg_\theta$ in the norm $\Vert a \Vert^2_\tau=\tau(a^* a)$.
The basic derivations extend to closed skew-adjoint operators on
$\Hilbert_\tau$. Denote by $\rz^{n,k}$ the vector space $\rz^n$
endowed with the indefinite metric $q_{n,k}$ and let
$\Cl_{n,k}^c$ be the corresponding Clifford algebra. Let
$\Delta_{n,k}$ be the natural Clifford module for $\Cl_{n,k}^c$.
Denote by $\gamma(v)$ the representation of $\rz^n \subset \Cl_{n,k}^c$ on
$\Delta_{n,k}$.
We choose a basis $\{e_i\}$ in $\rz^n$ such that the $\gamma_i :=
\gamma(e_i)$ satisfy
$\gamma_i^2=-1$ for $i=1,\ldots,k$ and $\gamma_i^2=+1$ for $i=k+1,\ldots,n$.
We have
\begin{pro}
 Let $\Hilbert=\Hilbert_\tau \otimes \Delta_{n,k}$ and let
 $D$ be the closure of the operator
 \begin{gather}
  D_0:=\I^{k-1}\left( \sum_{i=1}^n \gamma_i \delta_i \right)
 \end{gather}
 on $\Hilbert$ with domain $\dom(D_0)=\alg_\theta \otimes \Delta_{n,k}$.
 Then $\Hilbert$ is a Krein space with the indefinite inner
 product defined by
 \begin{gather}
 (\psi_1 \otimes v_1,\psi_2 \otimes v_2):=\langle \psi_1,\psi_2
 \rangle_{\Hilbert_\tau} (v_1,v_2)_{\Delta_{n,k}},
 \end{gather}
 and $(\alg_\theta,\Hilbert,D)$ is a
 semi-Riemannian spectral triple. If $n$ is even the triple
 $(\alg_\theta,\Hilbert,D)$ is even.
\end{pro}
\begin{proof}
 Since $\Delta_{n,k}$ is finite dimensional and decomposable, each
 decomposition of $\Delta_{n,k}=V^+ \oplus V^-$ into positive and negative definite
 subspaces gives rise to a decomposition $\Hilbert=\Hilbert_\tau \otimes V^+
 \oplus \Hilbert_\tau \otimes V^-$. Clearly, the subspaces are intrinsically
 complete. Therefore, $\Hilbert$ is a Krein space. Next we show that $D_0$
 is essentially Krein-selfadjoint on $\Hilbert$. Clearly,
 $\J:=\I^{\frac{k(k+1)}{2}} \gamma_1 \cdots \gamma_k$ is a fundamental
 symmetry of $\Hilbert$ and it is enough to show that the symmetric operator
 $\J D_0$ is essentially selfadjoint on $\Hilbert$ endowed with the scalar
 product induced by $\J$.
 The vectors $u(y) \in \alg_\theta \subset \Hilbert_\tau$
 form a total set in $\Hilbert_\tau$ and it is easy to see that the vectors
 of the form $u(y) \otimes \psi$ are analytic for $\J D_0$. By Nelsons theorem
 $\J D_0$ is essentially selfadjoint on $\alg_\theta \otimes \Delta_{n,k} \subset
 \Hilbert$ and therefore, $D_0$ is essentially Krein-selfadjoint.
 For all $a \in \alg_\theta$ we have $[D,a]=\I^{k-1}\sum_i \gamma_i(\delta_i a)$,
 which is clearly a bounded operator. Hence, $(\alg_\theta,\Hilbert,D)$
 is a semi Riemannian spectral triple. For even $n$ this triple is even
 and the grading operator $\chi$ is just the grading operator in $\Cl_{n,k}^c$
 acting on the second tensor factor. In case $n$ is odd the triple is odd
 and we set $\chi=1$.
\end{proof}

In the following we will need the image of the universal differential forms
and the junk forms under the representation $\pi: \Omega \alg_\theta \to
\mathcal{B}(\Hilbert)$
associated with $(\alg_\theta,\Hilbert,D)$.
\begin{lem}\label{hilf}
 For the above defined semi-Riemannian spectral triple we have
 \begin{gather}
  \pi(\Omega^m \alg_\theta)=\left\{ \sum_j a^j \gamma(v_1^j)\cdots \gamma(v_m^j)\; ;\; a^j
  \in \alg_\theta, v_i^j \in \Delta_{n,k} \right\}, \\
  \pi(dj_0 \cap \Omega^m \alg_\theta)=\left\{ \sum_j a^j \gamma(v_1^j)\cdots \gamma(v_{m-2}^j)\; ;\;
  a^j \in \alg_\theta, v_i^j \in \Delta_{n,k} \right\}.
 \end{gather}
\end{lem}
\begin{proof}
 The first equation follows from the relation $[D,a]=\I^{k-1}\sum_i \gamma_i
 \delta_i(a)$. It remains to show that the second equation holds.
 Let $\omega$ be the $(m-1)$-form $(f_0 df_0 -df_0 f_0)df_1
 \cdots df_{m-2}$ with $f_0 = u_l$. We have $\delta_i f_0 = 2 \pi \I
 \delta_{il} f_0$. A short calculation shows that $\pi(\omega)=0$ and therefore,
 the form
 \begin{gather}
  \pi(d \omega)=- 8 \pi^2 \gamma_l^2 f^2_0 [D,f_1] \cdots [D,f_m]
 \end{gather}
 is an element of $\pi(dj_0 \cap \Omega^m \alg_\theta)$.
 The $\alg_\theta$-module generated by this form is the set $A^m$
 of elements of the form
 $\sum_j a^j \gamma(v_1^j) \cdots \gamma(v_{m-2}^j)$ with $a^j \in \alg_\theta$ and
 $v_1^j,\ldots,v_{k-2}^j \in \Delta_{n,k}$. Therefore, $A^m \subset \pi(dj_0 \cap
 \Omega^m\alg_\theta)$. In case $m-1 \geq n$ this shows that
 $\pi(\Omega^m \alg_\theta) = \pi(dj_0 \cap \Omega^m\alg_\theta)$ and the
 above formula is a consequence of this. We treat the case $m \leq n$.
 Suppose that $\omega=\sum_j f_0^j df_1^j \cdots df_{m-1}^j$
 and that
 \begin{gather}
  \pi(\omega)=\gamma_{\mu_1}\cdots\gamma_{\mu_{m-1}}
  \sum_j f_0^j \delta_{\mu_1}f_1^j \cdots \delta_{\mu_{m-1}}f_{m-1}^j=0.
 \end{gather}
 This implies that
 \begin{gather}
  \sum_j f_0^j \delta_{[\mu_1}f_1^j \cdots \delta_{\mu_{m-1}]}f_{m-1}^j=0,
 \end{gather}
 where the square bracket indicates the complete anti-symmetrization of the
 indices. If we apply $\delta_{\mu_0}$ to the left of this equation
 and anti-symmetrize in all indices we obtain
 \begin{gather}
  \sum_j \delta_{[\mu_0}f_0^j \delta_{\mu_1}f_1^j \cdots \delta_{\mu_{m-1}]}f_{m-1}^j=0.
 \end{gather}
 Since
 \begin{gather}
  \pi(d \omega) = \gamma_{\mu_0}\cdots\gamma_{\mu_{m-1}}
  \sum_j \delta_{\mu_0}f_0^j \delta_{\mu_1}f_1^j \cdots \delta_{\mu_{m-1}}f_{m-1}^j,
 \end{gather}
 we finally obtain $\pi(d \omega) \in A^m$.
\end{proof}

The above lemma implies that $\Omega_D\alg_\theta \cong \bigoplus_{m} \alg_\theta \otimes \Lambda^m
\rz^n$ and the differential is given by
$d(a \otimes e_1 \wedge \ldots \wedge e_k)= \sum_i \delta_i(a) \otimes e_i \wedge e_1 \wedge
\ldots \wedge e_k $. Here $e_i$ is a distinguished basis in $\rz^n$.

Each spacelike reflection in $\rz^{n,k}$ induces a fundamental symmetry
$\tilde \J_r$ of $\Delta_{n,k}$ and clearly, $\J_r:=\textrm{id} \otimes \tilde \J_r$
is a fundamental symmetry of $\Hilbert$.
All these fundamental symmetries are in fact equivalent and hence induce
the same smooth structure on $(\alg_\theta,\Hilbert,D)$.
\begin{pro}\label{iso}
 Let $r_1$ and $r_2$ be two spacelike reflections of $\rz^{n,k}$.
 Then $\J_{r_1} \sim \J_{r_2}$,
 i.e. $\dom(\Delta_{\J_{r_1}}^s)$ and $\dom(\Delta_{\J_{r_2}}^s)$
 are equal and carry the same topology for all $s \in \rz$.
 Moreover $\Hilbert^\infty=\alg_\tau \otimes \Delta_{n,k}$.
\end{pro}
\begin{proof}
 Let $E$ be the spinor bundle on the commutative torus $T^{n,k}$
 with the flat semi-Riemannian metric of signature $(n,k)$.
 Let $\Hilbert_c$ be the Krein space of square integrable sections
 of $E$. The map $W : \Hilbert_\tau \to L^2(T^{n,k})$ defined by
 $W(u(y))=e^{2 \pi i (y,x)}$ is unitary and satisfies $W \delta_i W^{-1} = \partial_i$.
 The map $U:=W \otimes \textrm{id}$ is an isometric isomorphism of the Krein spaces
 $\Hilbert$ and $\Hilbert_c$ and $U D U^{-1}$ coincides with
 $\I^{k}\Dirac$, where $\Dirac$ is
 the Dirac operator on the torus. Furthermore $U \J_r U^{-1}$ are admissible
 fundamental symmetries of the canonical spectral triple associated with $T^{n,k}$.
 As a consequence the operators
 $U \Delta_{\J_r} U^{-1}$ are classical pseudo\-differential operators
 of first order and hence, $\dom(\Delta_{\J_r}^s)=U^{-1} H_s(E)$ for $s>0$
 where $H_s(E)$ is the space of Sobolev sections of order $s$ of $E$.
 Therefore, $\J_{r_1} \sim \J_{r_2}$. The equation
 $\Hilbert^\infty=\alg_\tau \otimes \Delta_{n,k}$ follows from
 $W^{-1} C^\infty(T^{n,k}) = \alg_\theta \subset \Hilbert_\tau$, which is
 easy to check.
\end{proof}

We view in the following $(\alg_\theta,\Hilbert,D)$ as a smooth
semi-Riemannian spectral triple with the above defined smooth structure
and refer to it as the noncommutative semi-Riemannian torus $T^{n,k}_\theta$.
For simplicity we restrict our considerations to the case where
the algebra $\alg_\theta$ has trivial center.

\begin{theorem}
 Suppose that $\alg_\theta$ has trivial center. Then the set of admissible
 fundamental symmetries of $(\alg_\theta,\Hilbert,D)$ coincides with
 the set
 $$\{ \J_r \;; r \textrm{ is a spacelike reflection of } \Delta_{n,k} \}.$$
\end{theorem}
\begin{proof}
 Let $r$ be a spacelike reflection of $\Delta_{n,k}$. By construction $\J_r$ is smooth.
 We first show that $\J_r$ is admissible. Clearly, $\J_r$ commutes with
 all elements of $\alg_\theta$ and $\J \chi \J = \chi^+$. Lemma \ref{hilf} shows that indeed
 $\J \pi(j \cap \Omega^p \alg) \J = \pi(j \cap \Omega^p \alg)$
 and $\J \pi(\Omega \alg) \J = \pi(\Omega \alg)$. Therefore, $\J_r$
 is admissible.
 Now suppose conversely that $\J$ is an admissible fundamental symmetry.
 Since $\J$ commutes with $\alg_\theta$, we can view $\J$ as an element
 in $\alg_\theta' \otimes \textrm{End}(\Delta_{n,k})$, where
 $\alg_\theta'$ is the commutant of $\alg_\theta$ in
 $\mathcal{B}(\Hilbert_\tau)$. Since $\J$ is smooth, it is even
 an element of $\alg_\theta^{\textrm{opp}} \otimes \textrm{End}(\Delta_{n,k})$,
 where $\alg_\theta^{\textrm{opp}}$ denotes the opposite algebra
 of $\alg_\theta$ which acts on $\Hilbert_\tau$ from the right.
 The space $\pi(\Omega^1\alg_\theta)$ is invariant under conjugation
 by $\J$. Therefore, the matrices $\J \gamma_i \J$ must commute with all elements
 of $\alg_\theta^{\textrm{opp}}$ and therefore have entries in the center of 
 $\alg_\theta^{\textrm{opp}}$, which is trivial. Hence, the vector
 space spanned by the $\gamma_i$ is invariant under conjugation by $\J$.
 In the same way as in the proof of Prop. \ref{scalar} one
 checks that the map $r: \rz^{n,k} \to \rz^{n,k}$
 defined by $\J \gamma(v) \J = (-1)^{k} \gamma(rv)$ is a spacelike reflection.
 Hence, there exists a spacelike reflection $r$ of
 $\Delta_{n,k}$ such that $\J \gamma_i \J = \J_r \gamma_i \J_r$.
 Denote by $a$ the operator $\J_r \J$. Then $a$ commutes with all $\gamma_i$
 and commutes with $\chi$. Hence, $a \in \alg_\theta^{\textrm{opp}}$
 and therefore, $a$ commutes with $\J_r$. We finally get from
 $a^+ a = a a^+=1$ the equality $a^2=1$. Since both $\J_r$ and $\J$
 give rise to positive scalar products, $a$ must be positive in the
 $\J_r$-scalar product and therefore, $a=1$. We conclude that $\J=\J_r$.
\end{proof}

\begin{theorem} \label{last}
 Suppose that $\alg_\theta$ has trivial center.
 The smooth semi-Riemannian spectral triple $(\alg_\theta,\Hilbert,D)$
 is $n^+$-summable and for all $a \in \alg_\theta$ and each admissible
 fundamental symmetry $\J$ we have
 \begin{gather}
  \Dix(a \Delta_\J^{-n})=\frac{1}{c(n)}\tau(a), \\
  \Dix(a D^2 \Delta_\J^{-n-2})=(-1)^{k}\frac{n-2k}{n} \Dix(a \Delta_\J^{-n}).
 \end{gather}
\end{theorem}

\begin{proof}
 Let $E$ be the spinor bundle on the commutative torus $T^{n,k}$
 with the flat semi-Riemannian metric of signature $(n,k)$.
 Let $\Hilbert_c$ be the Krein space of square integrable sections
 of $E$. In the proof of Prop. \ref{iso} we constructed an isomorphism
 of Krein spaces $U:\Hilbert \to \Hilbert_c$ such that
 $U D U^{-1}=\I^{k} \Dirac$, where $\Dirac$ is the 
 Dirac operator on $T^{n,k}$. Moreover the $U \J_r U^{-1}$
 are admissible fundamental symmetries of the canonical spectral
 triple associated with $T^{n,k}$. Therefore, by theorem \ref{th2}
 $\Dix(\Delta_{\J}^{-n})= c(n)^{-1}$
 and  $\Dix(D^2 \Delta_\J^{-n-2})=(-1)^{k}\frac{n-2k}{n} \Dix(\Delta_\J^{-n})$
 for all admissible fundamental symmetries.
 The proof is finished if we can show that 
 $\Dix(u(y) \Delta_\J^{-n})=0$ and $\Dix(u(y) D^2
 \Delta_\J^{-n-2})=0$ whenever $y \not= 0$. Let $\{\psi_i\}$ be an orthonormal basis in
 $\Delta_{n,k}$. Then the elements $\phi_{y,i}:=u(y) \otimes \psi_i \in \Hilbert_\tau$
 form an orthonormal basis in $\Hilbert$ and they are eigenvectors of
 $\Delta_\J^{-n}$ and of $D^2$. By Lemma 7.17 in \cite{Gracia-Bondia:2001tr}
 we have $\Dix(u(y)\Delta_\J^{-n})=\lim_{p\to \infty} \Tr(E_p u(y)\Delta_\J^{-n})$,
 where $E_p$ is the orthogonal projector onto the subspace generated
 by the first $n$ eigenvectors of $\Delta_\J^{-1}$ and whenever the limit
 exists.
 But since $\langle \phi_{y,i}, u(y') \phi_{y,i} \rangle_\J=0$ for all $y' \not= 0$,
 we get $\Tr(E_p u(y')\Delta_\J^{-n})=0$ and consequently
 $\Dix(u(y')\Delta_\J^{-n})=0$.
 The same argument gives $\Dix(u(y') D^2 \Delta_\J^{-n-2})=0$
\end{proof}

\section{Outlook}
We showed that it is possible to extract the dimension, the signature
and a notion of integration from the spectral data of a semi-Riemannian
manifold. It would certainly be interesting if one could obtain the
Einstein-Hilbert action in a similar way as in the Riemannian case
(see \cite{Kastler:1995fg, Kalau:1995de}). This can probably not
be done straightforwardly, but may require some averaging of
expressions of the form $\mathrm{Wres(D^2 \Delta_{\J}^{-n})}$ over the set
of admissible fundamental symmetries.

Another interesting question is, which further conditions on the admissible
fundamental symmetries are necessary in the general situation to
guarantee that the functionals $\Dix(\;\cdot\; \Delta_{\J}^{-n})$ and
$\Dix(\;\cdot\; D^2 \Delta_{\J}^{-n-2})$
on the algebra generated by $\alg$ and $[D,\alg]$ do not depend on the choice
of $\J$.

As far as the noncommutative tori are concerned we believe that an analogue of
Theorem \ref{last} holds in case the center of $\alg_\theta$
is not trivial. One should be able to proof this in a similar way as we did
it here for the case of a trivial center.

\section{Acknowledgement}

The author would like to thank H. Baum, H. Grosse, R. Nest, C.W. Rupp and
M. Wollenberg for useful discussions and comments.
This work was supported by the
Deutsche Forschungsgemeinschaft within the scope of the postgraduate
scholarship programme ``Graduiertenkolleg Quantenfeldtheorie'' at the
University of Leipzig.

\end{document}